\documentclass[a4paper,english,fleqn,leqno]{llncs}
\usepackage{xspace}
\newcommand{\module}{course\xspace}
\newcommand{\modules}{courses\xspace}

\newcommand{\AC}{\textsc{ac}\xspace}
\newcommand{\MR}{\textsc{mr}\xspace}
\newcommand{\JD}{\textsc{jd}\xspace}
\newcommand{\CD}{\textsc{cd}\xspace}
\newcommand{\MF}{\textsc{mf}\xspace}

\newcommand{\PK}{\textsc{pk}\xspace}
\newcommand{\SK}{\textsc{sk}\xspace}
\newcommand{\PO}{\textsc{po}\xspace}

\newcommand{\NS}{\textsc{ns}\xspace}
\newcommand{\RZ}{\textsc{rz}\xspace}

\usepackage{hyperref}
\hypersetup{
    linkcolor=blue,
    filecolor=magenta,
}

\usepackage[capitalise,nameinlink]{cleveref}
\title{Rooting Formal Methods  \\
within Higher Education Curricula \\  for Computer Science and Software Engineering \\
--- A White Paper ---
}

\author{
  Antonio Cerone \inst{1}\and
  Markus Roggenbach\inst{2} \and
  James Davenport\inst{3} \and
  Casey Denner \inst{4} \and
  Marie Farrell \inst{5} \and
  Magne Haveraaen \inst{6} \and
  Faron Moller \inst {4} \and
  Philipp Körner \inst{7} \and
  Sebastian Krings \inst{8}\and
  Peter \"{O}lveczky \inst{9} \and
  Bernd-Holger Schlingloff \inst{10} \and
  Nikolay Shilov\inst{11} \and
  Rustam Zhumagambetov \inst{12}
}

\institute{
  Nazarbayev University, Nur-Sultan, Kazakhstan \\   
  \email{antonio.cerone@nu.edu.kz}\\[.5ex] \and
  Swansea University, United Kindom\\                
  \email{m.roggenbach@swansea.ac.uk}\\[1ex] \and        
  University of Bath, United Kingdom; \and               
  Swansea University, United Kingom; \and               
  University of Manchester, United Kingdom;\and        
  University of Bergen, Norway; \and                    
  Heinrich-Heine-Universit{\"a}t, Germany; \and         
  Niederrhein University of Applied Sciences,  Germany; \and  
  University of Oslo, Norway;  \and                   
  Humboldt-Universit{\"a}t zu Berlin, Germany;  \and    
  Innoplolis University, Russia; \and                   
  Nazarbayev University, Nur-Sultan, Kazakhstan      
}

\begin{document}

\maketitle

\begin{abstract}
This white paper argues that formal methods need to be better rooted
in higher education curricula for computer science and software
engineering programmes of study.
To this end, it advocates
\begin{itemize}
\item
improved teaching of formal methods;
\item
systematic highlighting of formal methods within existing, `classical'
computer science \modules; and
\item
the inclusion of a compulsory formal methods \module
in computer science
and software engineering curricula.
\end{itemize}

\smallskip

These recommendations are based on the observations that
\begin{itemize}
\item
formal methods are an essential and cost-effective means to increase
software quality; however
\item
computer science and software engineering programmes
typically fail to provide adequate training in formal methods;
and thus
\item
there is a lack of computer science graduates who are
qualified to apply formal methods in industry.
\end{itemize}

\smallskip

This white paper is the result of a collective effort by authors and
participants of the \textit{1st International Workshop on
Formal Methods -- Fun for Everybody}
which was held in Bergen, Norway, 2-3 December
2019. As such, it represents insights based on learning and teaching
computer science and software engineering (with or without formal
methods) at various universities across Europe.
\end{abstract}







\section{Introduction}

The greatest contribution that universities make to industrial
practices is through releasing legions of graduates every year.
When properly equipped with a scholarly education,
these graduates challenge established processes and pave the way for new
approaches.
In the increasingly-digital world we live in,
the scope for this is arguably greatest
in the software industry, particularly given that
the public perception -- and indeed the reality -- is that
software is inherently unreliable.

Advances in digital technology take place at an astronomical rate,
unfettered by regulations which would hinder progress
in other scientific endeavours.
There are generally few established principles in place
to ensure that new software systems are as reliable as,
say, 
a new vaccine.
Software engineers demonstrate success in their company by
releasing systems which, for \emph{almost} all intents and purposes,
\emph{appear} to work.
Because of the benefits these advances offer society,
the public are generally accepting of
-- and, indeed, used to --
software failures.

This situation persists in spite of the fact that
computer science and software engineering research has developed
a multitude of design principles which could
help to improve software quality~\cite{barnes11}.
It has been over half a century since Robert Floyd's
seminal paper~\cite{Floyd67} set out the means by which
computer programs could be analysed to determine their functional correctness,
and formal methods for developing correct software
have been steadily devised and refined ever since.
The typical computer science or software engineering graduate,
however, leaves university with little or no knowledge of formal methods,
and even a dislike for whatever formal methods
they have encountered in their studies.
Thus, rather than opening doors for formal methods
in (software) industry, university education seems
to have a detrimental effect.

Due to their ubiquity, software failures are overlooked by society
as they tend to result in nothing more serious than
delays and frustrations. We accept as mere inconvenience
when a software failure results in a delayed train or
an out-of-order cash machine or a need to repeatedly enter
details into a website.
However, the problems of systems failures become
more serious (costly, deadly, invasive) as automatic control systems
find their way into virtually every aspect of our daily lives.
This increasing reliance on computer systems makes it essential
to develop and maintain software in which the possibility,
and probability, of hazardous errors is minimised.
Formal methods offer cost-efficient means to achieve
the required high degree of software quality.

A major reason that students (and, in turn, software engineers)
have a negative attitude towards formal methods is
that these are not introduced with due care
during the early stages of higher education.
Left to the theoretical computer science professor,
such courses often start with fearful terms like state machine,
logical inference, mathematical semantics, etc.,
without providing elementary explanations of the basic notions
which relate these to the practice of software development.
In their defence, formal methods professors often find it difficult to deliver
the subject due to students' scepticism~\cite{Zhumagambetov2020},
which arises from the generally limited or non-existent exposure
to formal methods in the rest of the curriculum. Boute~\cite{83} and
Sekerinski~\cite{822} observe that limited references from other
subjects and isolated use are the main factors leading to students'
low opinion. Even worse, students perceive formal methods to be
unsuitable for actual software engineering~\cite{84} or even an
``additional burden''~\cite{82}.

In this white paper we
analyse what hinders a successful formal methods education, and make
constructive suggestions about how to change the situation. We are
convinced that such changes are a prerequisite for formal methods to
become widely accepted in industry.
We analyse the current situation of formal methods
teaching and explore ways which we think will be engaging for students
and practitioners alike. Our vision is that formal methods can be
taught in such a way that both students and lecturers will enjoy
formal methods teaching.

This white paper is the result of a collective effort by authors and
participants at the 1st International Workshop ``Formal Methods – Fun
for Everybody'', which was held in Bergen, Norway, 2-3 December
2019. At the workshop, there were several discussion sessions. Based
on these, the two lead authors devised a paper outline, which was
subsequently ``populated'' with text snippets written by all
authors. The resulting draft was carefully edited, and agreed upon by
all authors. By its very nature, this white paper offers a spectrum of
opinions, in particular in the personal statements. What unites us are
the following beliefs:
\begin{itemize}
\item
Current software engineering practices fail to deliver dependable
software.
\item
Formal methods are capable of improving this situation, and are
beneficial and cost-effective for mainstream software development.

\item
Education in formal methods is key to progress things.
\item
Education in formal methods needs to be transformed.
\end{itemize}

In \cref{sec:challenges}, we analyse the challenges in teaching formal
methods. In \cref{sec:fun}, we collect ideas about how to teach formal
methods -- the fun way. In \cref{sec:visibility}, we discuss how to
increase the visibility of formal methods throughout the
curriculum. In \cref{sec:syllabus}, we suggest a syllabus for a
compulsory formal methods \module. Finally, we discuss how to assess
such teaching efforts in \cref{sec:efforts},
before making concluding remarks in \cref{sec:conclusion}.

\section{Challenges in teaching formal methods}\label{sec:challenges}

Teaching of formal methods faces a number of challenges. Currently, as
a knowledge area, formal methods are virtually absent from curricula
in computer science or software engineering.  Formal Methods barely
appear in the ACM/IEEE 2014 Software Engineering Curriculum, and
indeed the development of formal specifications is explicitly deemed
to be inappropriate for a capstone project~\cite[p.~56]{acmSE}. Moreover, many
students have an incorrect perception of what formal methods are
about. Formal methods neither make the headlines nor are a popular
topic in social networks, nor are they visibly used by industry. It is
also the case that colleagues as well as students have misguided ideas
concerning the mathematical background required to utilise formal
methods. In the following, we elaborate on these topics. The section
concludes with personal statements.

We begin our discussion by providing a working definition,
cf.~\cite{0815}, of what a formal method might be.
\begin{definition}
A formal method $M$ can be seen to consist of the three elements
syntax, semantics, and method:
\begin{itemize}
\item Syntax: the precise description of the
form of objects (strings or graphs) belonging to $M$.
\item
Semantics:
the `meaning' of the syntactic objects of $M$, in general by a
mapping into some mathematical structure.
\item
Method: algorithmic
ways of transforming syntactic objects, in order to gain some insight
about them.
\end{itemize}
\end{definition}
A typical example of a formal method is the process algebra
CSP: its syntax is given in form of a grammar; there are various
formal semantics (operational, denotational, and axiomatic ones); and
there are proof methods for refinement via model checking and theorem
proving.

Applying this definition, e.g., to the programming language Pascal, we
see that it also qualifies as a formal method. It has a defined syntax
and formal semantics; and each compiler and static analyser provides a
method, the Hoare calculus would be another instance of a method.

UML on the other hand does not qualify as a formal method. The syntax is
largely fixed via meta models, and there are various methods
available, e.g., for code generation (e.g., from class diagrams or
state machines). However, proposed semantics for UML contain several critical
``variation points'' and has -- to the best of our knowledge -- never
been fully formalised.

\subsection{On the absence of formal methods from computer science and software engineering curricula}
\label{sect-challenges-absence}

Anecdotal evidence suggests that current computer science and software
engineering curricula rarely cover formal methods to a large extent.
We exemplify this observation by providing an historic
perspective on programming education, an element central to all
curricula.

In the late 1980s, Pascal was a dominant teaching language for
beginning programming students. Pascal is a small, structured
programming language with a syntax designed to be easy to
parse~\cite{iso_pascal1990}. Most textbooks of the time presented the Pascal
language using syntax diagrams, alerting the students to the idea of
context free grammars, e.g.,~\cite{cooperclancy1982pascal}. The
element of syntax was taught as an integral part of programming.
Some textbooks included
the entire ISO Pascal standard, thus making the
students aware of language definition documents. 
\footnote{Pattis \cite{10.1145/191029.191155} even suggested teaching
  Extended Backus-Naur Form (EBNF) as the first topic in computer
  science. Not to facilitate presenting the syntax of a programming
  language, but because EBNF is a microcosm of programming. With no
  prerequisites, students are introduced to a variety of fundamental
  concepts in programming: formal systems, abstraction, control
  structures, equivalence of descriptions, the difference between
  syntax and semantics, and the relative power of recursion versus
  iteration.}  For those specifically interested, Pascal had a widely
available formal semantics~\cite{HoareW73/DBLP:journals/acta}. Robust
programming, i.e., checking preconditions, was an essential part of
programming courses. Some universities would even have space for a
formal methods course, typically based on Hoare logic, in their
undergraduate curriculum: i.e., a formal method was taught.


About 20 years ago, Pascal was superseded by Java as the dominating
teaching language. Java is a much more complex language than Pascal;
it supports object-oriented development, and it has large support
libraries. Thus, in the transition to Java, precise syntax and
semantics was replaced by a more example-driven approach,
e.g.,~\cite{deiteldeitel2007java}, where the first half contains
similar material to~\cite{cooperclancy1982pascal}. Verification tools
such as Java
Pathfinder\footnote{\url{https://github.com/javapathfinder/jpf-core/wiki}}
rarely made it into the syllabus of a programming course. Instead,
students needed to learn more methodology, such as object-orientation,
test-driven design and agile methods. All of this reduces the
students' exposure to formality, such as formal syntax or precise
semantics\footnote{The recent \emph{The Java® Language Specification,
    Java SE 14 Edition} is 800
  pages~\cite{GoslingJoySteeleBrachaBuckleySmithBierman2020jsr_389-java_reference}
  and not easily digestible.}, making the gap to formal methods
larger. Further, the pragmatics of software development take up more
of the curriculum, leaving less space for a formal methods course in
the core curriculum. Dewar and Schonberg support this critical
assessment: ``It is our view that Computer Science education is
neglecting basic skills, in particular in the areas of programming and
formal methods. We consider that the general adoption of Java as a
first programming language is in part responsible for this decline.''
\cite{crosstalk}

In recent years, Python has emerged into a common teaching language for
programming. The move towards Python represents a change back to a much
smaller language than Java. The Python reference document is just 160
pages, and its formal grammar is only four
pages~\cite{Rossumetal2020python_reference}. This should make it possible to
at least expose the students to formal syntax and a standardisation
document. However, the typing and semantic model of Python remains
complex, and is not easily formalised.

Thus, while current programming education based on Java often fails to
provide foundations for formal methods by discussing syntax and
semantics, the move towards Python provides the silver lining that the
element of syntax might again become a part of standard education in
programming.


\subsection{Students' perception of formal methods}

The reduced exposure to formal approaches, as described in
Section~\ref{sect-challenges-absence}, supports university students'
misconception that formal methods are a difficult topic with little or
no practical relevance. This keeps students away from formal methods
during their undergraduate studies. Even worse, it leads them to
embrace the common belief that mathematics and computer science are
two independent, fully distinct disciplines. Computer science is
rather identified with programming, which, in turn, is seen more like
an art rather than a scientific activity~\cite{CeroneLermer2020}.
Interestingly, this view has even been supported not only by the
pragmatic evolution of programming languages outlined in the previous
paragraphs, but also by some academic publications claiming that
rigorous mathematical knowledge is not necessary for computer science
practitioners~\cite{Glass2000}. Finally, this view has been
paradoxically encouraged by the introduction of computer science in
high schools. In fact, although in several schools computer science
has been introduced as a stand-alone subject, it is not connected with
mathematics but, instead, it is presented as a `service subject'
intrinsically tied to the use of computers. Scope of the subject is to
provide tools that facilitate students in carrying out their homework
and class projects~\cite{Cerone2020,Gibson2008}.



Although we can say that, on average, a typical computer science
student tends to have a negative perception of formal methods, in
reality lecturers observe a lot of variation between students, as well
as changes of perceptions in one direction or the other. Variations
in students can be observed starting from the first programming
courses. A slightly exaggerated categorisation goes as follows. On the
one hand, there are students who tackle programming in a purely
`artistic way' by sitting down at the computer and writing code
immediately, using debugging rather than problem solving to reach the
solution. On the other hand, there are students who start analysing
the problem using pen and paper, then draw diagrams, possibly write
pseudo-code, test their solution on paper and, only when they are
confident in their solution, they sit in front of a computer and
convert their solution into a program. Obviously, it is the latter
approach what lectures suggest. Normally, the former group of students
tend to have a negative perception of formal methods, whereas the
latter group tend to have a positive one. This partition of the
students in two groups appears more evident once recursion is
introduced in the programming course. The former group of students
will tend to hate recursion, the latter group will tend to love it.

These two opposite perceptions obviously occur in several
degrees. Moreover, they are not static but, at least potentially,
dynamic and may be either encouraged or hindered in various ways
throughout the course of undergraduate studies. The common
absence of formal semantics among the topics of programming courses
definitely keeps students away from an early exposure to formal
methods and prevents them from really understanding what formal
methods are. Being exposed to some basic operational semantics could
actually help students to better understand conditional and iterative
constructs, which are normally serious challenges for first year
students. Furthermore, recursion could be better understood, thus
providing the basis for a future interest in formal methods.

Concerning senior students, although for some of them their perception
of formal methods may have been strongly oriented towards the negative
side, there is hope to shift them towards the positive side. Senior
students tend to be very pragmatic and their minds are dominated by
the goal of entering the job market and the industrial
world. Therefore they will build a positive perception of formal
methods when presented with their pragmatic and industry-oriented aspects.

\subsection{Limited visibility of formal methods in media and industry}

How students perceive a knowledge area has many drivers, such as
personal success, like/dislike of certain academic teachers, their
grades, etc.  But maybe `coolness' is the dominant factor. During their
studies, students want to do something cool, maybe work with
AlphaZero\footnote{AlphaZero is the descendant of AlphaGo, the
  AI\ that became known for defeating Lee Sedol, the world's best Go
  player, in March of 2016.} or participate in a hackathon such as
Google's Hash Code. Students also strive to get `cool jobs', e.g.,
with Google, Facebook, Amazon, and the like. Currently, what one might
want to call the `coolness factor' of formal methods is rather
low. Formal methods make neither the headlines nor are prominent in
social media, nor are they visibly used by industry.

Besides studying, quite a number of students work on the side for
companies. In these jobs, students often see only small parts of the
overall job profile of a professional computer scientist or software
engineer. Many of these side jobs deal with having a quick and dirty
solution for some pressing problem, adapting software according to
customer requests, or building prototypes in order to find out whether
some concept works out. In contrast, mature students, coming back from
industry and getting into university education again, know about the
importance of quality assurance. But as they usually were not exposed
to formal methods in their jobs, they are often reluctant to study
them.

Luckily, there is some serious uptake of formal methods in
industry. The classic case of a safety-critical industry is railway
signalling, as described e.g.\ in~\cite{Fantechietal2013a}.  Ligne 14
of the Paris M\'etro had software built using the
B~method~\cite{Fantechietal2013a} and has now run for over 20 years without a
bug being reported. The ``High Integrity Systems'' unit of Altran
develops systems for, e.g., the railway signalling industry and air
traffic control, as well as tools and methodologies, such as the SPARK
subset of Ada~\cite{McCormickChapin2015a}. SPARK 2014 uses contracts
to describe the specification of components in a form that is suitable
for both static and dynamic verification.

Outside the safety-critical industry, a few `enlightened', large
information technology companies are beginning to use formal
methods:
\begin{itemize}
\item Google is developing an ecosystem for formal analysis
  tools~\cite{7194609}.
\item Facebook uses ``advanced static analysis'' as
  described in~\cite{DiStefanoetal2019a}.
\item Amazon's use of formal methods is discussed
  in~\cite{Newcombeetal2015a,8880058}. There is a more technical
  description of one component in~\cite{Chudnovetal2018a}.
\end{itemize}
If we look at Facebook,~\cite{DiStefanoetal2019a} reports that, in
many cases, ``we have gravitated toward a `diff time' deployment,
where analyzers participate as bots in code review, making automatic
comments when an engineer submits a code modification''. For their
Infer tool, which has its origin in the separation logic work
of~\cite{Calcagnoetal2011a}, they aim ``for Infer to run in 15--20min on
a diff on average''.  \par Similarly, at Altran, an attempt to check
source code into the main repository (the equivalent of
\verb+git push+) generates a requirement to prove the appropriate
contracts, and the verification conditions that ensure, for example,
no numeric overflow. An important requirement here is that this
verification be ``reasonably fast''. \cite{BrainSchanda2012a}~describes
their work here as ``this changes the qualitative time band for a
large scale industrial project from `Nightly' to `Coffee'.''  Both
Facebook and Altran argue that the primary purpose of this time
requirement is to avoid `context switch' in the developer's brain.

Further changes could be initiated by
academics. ``Two-hundred-terabyte maths proof is largest ever''
reported Nature in May 2016\footnote{
  \href{https://www.nature.com/news/two-hundred-terabyte-maths-proof-is-largest-ever-1.19990}{Nature, 26
    May 2016.}} and wrote: ``Three computer scientists have announced
the largest-ever mathematics proof: a file that comes in at a whopping
200 terabytes, roughly equivalent to all the digitized text held by
the US Library of Congress. The researchers have created a 68-gigabyte
compressed version of their solution -- which would allow anyone with
about 30,000 hours of spare processor time to download, reconstruct
and verify it -- but a human could never hope to read through it.''
The results that triggered this media interest concerns the
Pythagorean Triples Problem.  ``We consider all partitions of the set
$\{1, 2, \dots \}$ of natural numbers into finitely many parts, and
the question is whether always at least one part contains a
Pythagorean triple $(a, b, c)$ with $a^2 + b^2 = c^2.$ For example
when splitting into odd and even numbers, then the odd part does not
contain a Pythagorean triple (due to odd plus odd = even), but the
even part contains for example $6^2 + 8^2 = 10^2.$ We show that the
answer is yes when partitioning into two parts, and we conjecture the
answer to be yes for any finite size of the
partition.''~\cite{DBLP:journals/cacm/HeuleK17} Such results
triggering media interest could possibly change the situation. Another
approach could be to organise, say, verification competitions at a
student level. They would need to provide a stimulating social
environment by being accessible to all students, and could be
supported by elements such as cool prizes and free pizza.

\subsection{Students' mathematical background}

The seeming need for a solid mathematical background is often an
argument against teaching formal methods. However, reflecting on the
three elements of a formal method, grasping the syntax of a formal
method is not more involved than understanding the syntax of a
programming language: both are given by grammars. Grammars for formal
methods are usually smaller than those for programming
languages. However, students learn programming languages by trial and
error on a computer, where the compiler/interpreter provides feedback
on syntax errors. As discussed in Section
\ref{sect-challenges-absence}, standard programming courses mostly
take an example-driven approach to syntax. In contrast, in formal
methods students are often presented with a grammar for the
syntax. For students, this often provides the first mathematical
hurdle\footnote{This is not eased by the often poor error messages
  provided by formal method tools.}. The challenge in formal methods
teaching therefore lies in adopting a more example-driven style when
it comes to syntax.

The semantics of a formal method is inherently mathematical in nature:
in logic it is given in terms of the satisfaction of a formula by a
model, process algebra utilizes structural operational semantics or
denotational semantics, etc.

However, in a basic \module focused upon the application of formal
methods, it would be enough to point out that such formal semantics
exists and to hint at its nature. The teaching challenge lies in
providing an explorative approach to semantics via tools. In logic,
this could follow ideas such as Tarski's world. In process algebra,
one can explore processes by simulating them. In such a set-up, students
could develop their own formal models and explore them, i.e., tools
provide students with a similar feedback like running a computer
program. Another idea would be to use a semantics compatible with the
programming languages students are using. For instance in axiom-based
testing, the `axioms' can be interpreted as code in the programming
language, thus utilising the students' programming background.

In an advanced \module, in addition to such an explorative approach,
the formal semantics itself needs to be presented. This will require a
good mathematical background from the students.

Finally, the method aspect of a formal method is best presented
through the use of a tool that automates the analysis in which one is
interested. Running a tool would not require any mathematical
background at all. Understanding the result of a method applied to a
concrete example is usually immediate. An advanced \module would
address the mathematical details of why a method is sound.

These considerations refute the common prejudice that teaching formal
methods requires students to have a profound mathematical background.
An explorative teaching approach can make formal methods accessible
even to students who like to program the `artistic way'. This is
supported by experience reports such as: ``Engineers from entry level
to principal have been able to learn TLA+ from scratch and get useful
results in two to three weeks''~\cite{Newcombeetal2015a}.

\subsection{Personal statements}

In the order in which they were contributed, we present a number of
personal statements by the co-authors.

\paragraph{\SK.}
One challenge in teaching formal methods is to spark an initial
interest. This is the case, because links are weak between formal methods
and the current hot topics in computer science. Many students steer
towards what currently is perceived to dominate the future: data
science and artificial intelligence, to name a just a few.

To overcome this, the formal methods community should strive to
demonstrate its relevance, beyond `classical' topics such as railway
engineering. Correctness is as relevant in the new, upcoming areas of
computer science as it is in the classical ones.

\paragraph{\PK.}
A similar thought adding to \SK: many students do not even have
a clear idea of what formal methods are! They have heard of other
areas such as machine learning, databases, operating systems, computer
networks, compiler construction, and have an idea what is going on
there. It's hard to encounter many aspects of formal methods in daily
programming life, especially for a student with a limited view. So, why
exactly would they pick a `no-name' course such as ``formal methods''
or ``model checking'' over the other choices?

\paragraph{\CD.}
The name of a \module makes a big difference: students tend to avoid
\modules that already sound complicated (i.e.\ anything math or
formal) in contrast to \modules that sound `useful' or `applicable' or
even just trendy. As a student, I had a \module named ``Modelling
Computer Systems'' that was on discrete mathematics. If it had been
called ``Discrete Mathematics'', I'm sure it would have put several
students on edge to begin with. Courses with names that contain tech
buzzwords may also sound more appealing to students, such as cyber
security, software testing, machine learning, artificial intelligence
etc. We should consider these trendy subjects and adjust formal
methods to be just as appealing, even if it means slightly adjusting
\module names.

\paragraph{\MF.}
The lack of reliable tools that are suitable for teaching formal
methods, as well as are scalable enough to demonstrate interesting and
realistic use cases, creates a barrier for students. Throughout our
course, we used a number of freely available formal methods and
students struggled to understand the error messages and other feedback
from the tools~\cite{Farrell2020}. This kind of ambiguous feedback causes
the students to lose interest and prevents them from engaging with the
tools in a positive, constructive way.  Furthermore, this usability
issue also hinders the uptake of these tools in industry. This is
somewhat of a vicious circle. Admittedly, most formal method tools are
academic in nature and thus often are aimed at being good for
publication. Better error messages and the like are often not
prioritized that way. This causes the industrial uptake to miss, which
decreases the focus again.

\subsection{A student's personal statement}

\paragraph{\RZ.}

My first introduction to formal methods was during my second year
(right after introductory programming courses but before software
engineering) in the GPU computing course. We used Petri nets for
modelling the classic dining philosophers' problem. One of the
motivations for using them was to avoid software failures. By
providing a mathematical proof with Petri nets, so the professor
claimed, we would be on course for success. At that time formal
methods looked to me like an advanced technique in software
development and a usual practice. My illusions were shattered later
when another professor pointed out that it takes numerous assumptions
for formal methods to work in the real world, and that often these
assumptions do not apply.

\section{Teaching formal methods –- the fun way}\label{sec:fun}

In this section we collect a number of personal views and ideas on how
teaching formal methods can be done the fun way.  While some authors,
see, e.g.~\cite{Cerone2013TeachingFM}, have written systematic
accounts of the topic, here we present a number of personal statements
in the order in which they were contributed.


\paragraph{\MF.}

Games can be useful when it comes to teaching formal methods in the
initial stages. However, to adequately demonstrate the importance of
formal methods there must also be an emphasis on building and
verifying software and not just on solving a puzzle, as entertaining
as that may be. Of course, computer science students will find
enjoyment in building systems, otherwise they would not be studying
the subject. So, perhaps setting them the task of developing and
verifying a simple, but realistic, model of a system would also be
beneficial while encouraging them to have fun with formal methods. In
this setting, games would ideally be placed at the beginning of the
course as a light-weight and fun introduction.

\paragraph{\JD.}
It is often difficult to motivate formal methods. Most students will
not go into the construction of safety-critical systems, important
though they are. Also, the specialist safety-critical companies tend
to do their own training (though they would really like to have to do
less!). It is perhaps easier to motivate formal methods with more common
examples. The Chromium
Project\footnote{\url{https://www.chromium.org/Home/chromium-security/memory-safety}}
is one example of `mainstream' software, viz.\ browsers, and shows
that the Chromium team is moving `more formal'.

\paragraph{\SK.}
Usually, what makes any course interesting is the applications and the
transfer of knowledge from classroom to reality. However, most formal
method courses rely on examples that, while interesting, are far away
from what students can experience and experiment with. We often rely
on examples from industry and spend quite a lot of time explaining
what a particular model is supposed to achieve exactly. I feel this
often distracts students. Rather than focusing on what formal methods
have to offer, we get lost in technical details. This is not the case
with games, especially if considering well-known ones. Usually, the
rules are known and (mostly….) agreed upon already and we can focus on
how a formal method can help us to get them right in our application.

Again, I strongly believe we should get away from the purely
theoretical approach to teaching formal methods to beginners. At least
for me, the theoretical advances in formal methods have always been a
means to an end. In order to appreciate them, one has to
experience what it means to try and reach the same end without
them. This however falls short in programming education in
general. Students proceed from smallish group projects to other
smallish group projects, while only seldom have to experience larger
refactoring, legacy code, etc. In an environment like this, formal
methods are less useful. Let's teach our students what programming is
like in reality: 90\% of the work is reworking legacy code, fixing
bugs and trying to understand why things are or are not working -- by
accident, this is where formal approaches could shine as well.
Another aspect that could make a formal methods course interesting is
to involve students in formal methods research rather than formal
methods application. We used to teach formal methods by discussion
software issues first and then having students try to find automatic
ways to detect them, leading from simple static analysis ideas to
model checking. The course has been thoroughly documented, also
showing that the approach was highly motivating for students
students~\cite{Krings2019}.

Notably, students (at least on the masters level) are able and willing
to do `actual research' in an inquiry-based course, eventually leading
to publications~\cite{Petrasch2019}.
The inquiry- or research-based approach has taught students the internals
of model checkers and how they can be efficiently implemented for
prototypical languages.

\paragraph{\PK.}
Shriram Krishnamurthi had a great Keynote
at FM'19\footnote{\url{https://www.youtube.com/watch?v=UCwyOSHRBi0}}.
One of the main points to take away from that is that tools are a large
issue. If you hit students with a full-blown industrial tool, they get
frustrating error messages, because they have no idea what is going
wrong (as the tool is able to understand a larger part of, e.g., a
specification language than the student and raises errors related to
other concepts). While it is nice to see that such tools are used in
practice, they might be the wrong means to learn formal methods.

In Düsseldorf, our group has worked on an approach based on Jupyter
notebooks~\cite{10.1007/978-3-030-48077-6_19}.  It allows evaluation
  of smaller expressions or predicates without a state-based approach,
  so students can learn and experiment with the logical foundations of
  the
  language\footnote{\url{https://gitlab.cs.uni-duesseldorf.de/general/stups/prob2-jupyter-kernel/-/blob/master/notebooks/tutorials/prob\_solver\_intro.ipynb}}
where it is used to solve some logic puzzles). It can also be used to
interact with B machines, so errors in a specification can be
explained and documented in a nicer way (that you can replay). We
think that might resolve some of the issues in teaching (but probably
not all).

\paragraph{\CD.}
Games are important, maybe even essential in teaching formal methods
and making it fun. As a teacher of all ages from 8 years old to
university level, I have found games to be one of the best tools to
use when teaching. Students understand games and want to win them,
naturally. When you explain to students that there is a method in
which they are either guaranteed to win, or indeed a method in which
the second player cannot win, their interest levels peak! Students
rush home to play the games against their parents and show off their
new found ability.

Games can also be taught to most age groups. As said in our other
paper in this volume
``Appealing to their existing understanding of how the world works,
using puzzles as a medium, students can quickly become comfortable
using mathematical concepts such as labelled transition
systems''~\cite{Moller2020}.

We have had success in asking 11 year olds to draw labelled transition
systems. If we start teaching them sooner, this could act as a base
from which we can build upon to further their understanding later on.

\paragraph{\MF.}
In our experience~\cite{Farrell2020}
the students found it difficult to bridge the gap between the theory that
was taught during the \module (e.g. natural deduction proofs) and the
associated tool support used during the lab sessions (e.g. Coq). As a
result, I am inclined to agree with \SK above in that the
students need to see how these methods can work in reality rather than
focus too much (although it is important and should be covered at some
level) on the theory.

\paragraph{\AC.}
The use of tools provides a great potential for introducing fun
in teaching formal methods.
This is particularly true for simulation and model-checking tools,
whose emphasis is in giving ``life'' to formal specifications rather than
getting involved in the complexity of a formal proof, as it happens,
instead, for theorem-proving tools. Moreover, formal methods can be
applied to a large range of problems, basically any problem, well
beyond the domain of computer science. These give chances to teachers
to propose fun problems, such as classical mathematical puzzles as
well as popular games and even video games, and to learners to select
problems that are close to their personal and professional
interests~\cite{CeroneLermer2020}.
One effective approach consists of providing learners with
examples of formal methods descriptions of video games and inviting
them to create formal models of their favourite video games. More in
general, learners may be invited to define any problem they wish,
formally specify/model it and carry out analysis with the support of
tools. It is actually important to blur the distinction between
learner and instructor by letting the learners drive the choice of
exercises and use their creativity to identify and specify potential
problems and invent new games.
Blurring such a distinction will also contribute to instill in students
a level of self-confidence that can lead students
to carry out ``actual research''~\cite{Petrasch2019}
and to actively contribute to curriculum
development~\cite{Zhumagambetov2020}.

We can conclude our discussion on the teacher's view about fun by
saying that if motivation is the dimension that allows learners to
build up interest in formal methods, fun is actually the essential
dimension to keep learners continuously engaged, thus assuring the
retention and possibly increase of their interest over
time~\cite{CeroneLermer2020,Cerone2016,0815}.  However, it is
important that the fun occurs from the perspective of the student, not
the teacher, and, if it is associated with some form of competition,
this much effectively fosters motivation and does not cause
frustration.  In fact, nothing could be worse than ``fun degenerating
into frustration'', which could be the case when a game that is fun
for the teacher is actually too complex or uninteresting for the
students or, especially in the case of school children, if the outcome
of competition is interpreted by the student as a form of
assessment~\cite{Cerone2020}.

\paragraph{\NS.}

Fun, puzzles, games and entertainment in teaching are not the
unique ingredients needed to improve formal methods education (more
general – computer science and software engineering education).  All
these (and something else) are just ways to engage (undergraduate)
students with the learning, studying, comprehension and mastering of
formal methods using curiosity and amusement.  We believe that the
experience of individual educators and expertise of research groups in
the field of formal methods popularization deserves a positive
attitude from the computer science, software engineering and (even)
mathematics academic community and industry.

Another opportunity (just as an example) is a competitive spirit that
is so appropriate for young people (in particular – for students of
computer science and software engineering departments).  International
competitions between formal methods tools (e.g. automated theorem
provers and satisfiability solvers) are popular, useful and valuable
from the industrial and research perspectives, but not from the
undergraduate education perspective.  Unfortunately, competitions
especially designed for (undergraduate) students (like Collegiate
Programming Contest\footnote{\url{https://icpc.baylor.edu/}}) are
still not involved in the education process in general and in formal
methods education in particular.  We hope that competitions of this
kind may be used better for engaging students with theory of computer
science and formal methods in software engineering~\cite{Shilov2002}.

\paragraph{\PO.}
I also disagree with the `puzzle'/`games'/`card tricks' approach.  I
do not think that they show the usefulness and relevance of formal
methods. I also use small games (lots of them!) in my second-year
course, up to blackjack, but only as small ``toy examples'' to get to
know the modeling language and tool.  On the other hand, real
industrial applications, as others write here, are too large and
complex to include in beginner's formal methods courses.  A good
compromise that I use (and describe in my FMfun'19 paper) are seminal
systems/algorithms that are the cornerstones of different other
domains and, equally important, of today's large software systems. For
example, 2-phase-commit (while simple) and Paxos (less so) are still
key building blocks in today's distributed systems.  I include key
designs from other courses and beyond, like cryptographic protocols
(modeling and breaking NSKP), distributed transactions (2PC),
distributed mutual exclusion, distributed leader election, transport
protocols like TCP, ABP, sliding window, and so on.  This shows the
relevance of formal methods on many kinds of systems, and are small
enough to easily model and analyze using formal methods, but might
still give students (and other professors!) an idea of the usefulness
of formal methods.

One final problem with games/tricks: even if you learn how to apply
your formal method to model and analyze such games, can you then apply
your formal method to a real distributed system such as Paxos or a
cryptographic protocol?

I refer to my paper ``Teaching Formal Methods for Fun Using
Maude''~\cite{peter} in this volume for a lengthier exposition of how I
think formal methods should be taught at the undergraduate level.

\subsection{Summarizing the ideas}
It is obviously impossible to establish general criteria to make formal methods
teaching a fun activity.
Fun cannot be characterised in an objective way and can only naturally
emerge from the interaction between teachers and students.
In fact, the emergence of fun is affected by the personalities of individual
teachers and students as well as by the interaction context in which
such different personalities meet in the classroom collaborative environment.
Here, different criteria have been suggested and discussed, including:
\begin{itemize}
  \item games and puzzles may represent a light-weight and fun introduction
    to formal methods;
  \item there should be an emphasis on building and verifying software
    for simple, but realistic, systems;
  \item teaching should focus on demonstrating that tools work rather than on
    delivering too much theory;
  \item students are likely to enjoy undertaking actual research activities;
  \item students should be involved in curricula development.
\end{itemize}

There is a general view among the co-authors that games and puzzles
can be useful when it comes to teaching formal methods in the initial
stages and represent a light-weight and fun introduction (\MF, \CD, \AC).
It is important to note that
this view includes former formal methods students who became formal
methods teachers~\cite{Moller2020}.  Games may be also associated with
some form of competition (\AC, \NS), which may
be within-class (\AC) or in terms of participation at an
international context (\NS).  Games and puzzles are also a
great tool to start formal methods education early, even by teaching
to school level children, as young as 10--11 (\CD, \AC).
Competition can also be beneficial in the context of school
children, but should to carefully planned in order to avoid being
interpreted by the student as a form of assessment, which therefore
inhibits rather than motivates the students~\cite{Cerone2020}.

In addition, there must also be some emphasis on building and
verifying software (\MF).  However, such a connection with
reality should be established in the right form to keep in line with
the fun determined by the game-based approach.  In fact, giving
students the task of developing and verifying a simple, but realistic,
model of a system would be beneficial while encouraging them to have
fun with formal methods (\MF).  However, on the one hand,
realistic, industrial systems are often far away from what students
can experience and experiment with (\SK) and most
students will not go into the construction of safety-critical systems,
important though they are (\JD).  On the other hand, the
specialist safety-critical companies tend to do their own training
(\JD), which may provide a very different perspective from
what students learn in formal methods courses.  Moreover, focusing on
examples from industry is very time consuming and often involves heavy
technical details and, as a consequence, may be distractive rather
than beneficial (\SK).  Instead, it might be more
effective to motivate formal methods with more common, but still
realistic examples, such as the Chromium Project (\JD).

There is a general agreement among the co-authors that students need
to see how formal methods work in reality using tools rather than
focusing too much on the theory (\SK, \MF,
\PK, \AC, \PO).  However, making
students use industrial tools may result in heavy frustration.  While
it is nice to see that such tools are used in practice, they might be
the wrong means to learn formal methods (\PK).

An final aspect that could make a formal methods course interesting is to involve
     students in formal methods research rather than formal methods application (\SK).
     In fact, students' publication are often highly
    appreciated~\cite{Petrasch2019,Zhumagambetov2020}.

\section{Increasing visibility of formal methods throughout the curriculum}\label{sec:visibility}

In common computer science and software engineering curricula, formal
methods play a minor role. There are at most one or two specialized
\modules focusing on teaching formal methods. Often, these \modules are
only weakly linked to the rest of the curriculum.

Formal methods fail to link to the current hot topics in computer
science and software engineering, both in teaching and research. In
consequence, even students with considerable interest in software
engineering are drawn towards courses such as data science, machine
learning or artificial intelligence. However, now that artificial
intelligence and machine learning techniques find their way into
safety critical systems (such as autonomous cars), correctness
considerations become more important every day.

The `winner-takes-all' nature of today's software industry (where
essentially \emph{one} product/service in each category `wins' and
makes billions, and other solutions fade away, e.g., Facebook for
social media; Google for search engines, eBay for online auctions,
Zoom for online discussions/teaching/meetings) justifies an upfront
investment in system quality. We note that major firms like
Google~\cite{7194609}, Facebook~\cite{DiStefanoetal2019a}, and
Amazon~\cite{Newcombeetal2015a} are all doing this, but this has yet to feed
through to their hiring practices, or to students' perceptions of what
they need to get a job at these favoured employers.

In consequence, an ideal integration of formal methods into a computer
science or software engineering curriculum should first and foremost
strive to present formal methods as a quality assurance tool to be
used in other areas, be it embedded systems engineering or machine
learning. This first contact to formal methods would aim at teaching
usage scenarios as well as techniques and how they are to be deployed.

We believe that showing the benefit of formal methods by discussing
applications to other areas will achieve two goals. First, it ensures
code quality and system functionality are considered as
critical. Furthermore, this initial contact to formal methods might
spark an interest into their development and improvement. Both topics
could then be a part of dedicated courses in formal methods.

While such a `casual' approach would be ideal, it would require
colleagues to be willing and to be able to teach small units on formal
methods. This might be an unrealistic assumption. Organising `guest
sessions' from formal methods experts might be a way forward.

To gain an acceptance of having more formal methods visibility in
a university curriculum, we need to persuade first and foremost our
colleagues. Ultimately they decide whether/how/how much formal methods
a university curriculum could/must contain. There is huge competition
for places on a curriculum between the different specialties/fields.
At least the older colleagues may remember times when formal methods
were not too useful.

The 2013 ``Curriculum Guidelines for Undergraduate Degree Programs in
Computer Science''~\cite{acmCS} lists 18 ``Knowledge Areas''. In the
following, we make a number of suggestions for formal method units in
some of these areas:

\begin{description}
\item[AL-Algorithms and Complexity:]
     formal verification of algorithms; model checking algorithms. \\
\item[DS-Discrete Structures:]
   logic, modelling, semantic foundations of formal methods. \\
\item[HCI-Human-Computer Interaction:]
    mode confusion problems; formal analysis of user dialogs; cognitive models. \\
\item[IAS-Information Assurance and Security:]
   formal analysis of security protocols. \\
\item[IM-Information Management:]
  specifying and analyzing both the correctness and the performance of cloud storage systems.  \\
\item [NC-Networking and Communication:]
  protocol verification. \\
\item [OS-Operating Systems:]
  parallel modelling; scheduling. \\
\item [PBD-Platform-based Development:]
  formal model based development. \\
\item [PD-Parallel and Distributed Computing:]
  process calculi; Petri nets. \\
\item [PL-Programming Languages:] how to analyse software written in a
  specific programming paradigm; compiler correctness; semantics of
  programming languages; program correctness.
\end{description}

\section{Syllabus of a compulsory formal methods \module}\label{sec:syllabus}

Besides increasing the visibility of formal methods throughout all \modules
and also having specialised advanced \modules on formal methods, we
suggest that curricula for computer science and software engineering
should include a compulsory formal methods \module.

The target audience for such a compulsory formal methods \module would be
the complete cohort of computer science / software engineering
students in year 2 or year 3 of a 3-year BSc degree programme.

Due to the wealth of available formal methods, we refrain from
proposing a unified or `standard' syllabus. Local expertise in
specific formal methods and application domains should be taken into
account. Therefore, we rather capture the essence of an ideal \module in a
generic way:

\paragraph{Introduction.}
\begin{itemize}
\item
The role of formal methods in the context of software engineering,
see, e.g., Roggenbach et. al~\cite{0815}, Chapter 1, for a thorough
discussion, and Barnes~\cite{barnes11} for a comparative case study.
\item
Success stories of formal methods, see, e.g., Roggenbach
et. al~\cite{0815} for a compilation of such stories, another good
source is Section 1.3.4 of Garavel's report \cite{garavel};
\item
Relating formal methods to current trends in computer science,
such as machine learning, where one can use machine learning to
improve formal methods~\cite{Amrani2018MLF}, or, a nascent field but
one that is growing in importance and has already attracted the
attention of ISO in the draft TR 24029-2, the application of formal
methods to big data~\cite{Aalst2016,Camilli2014,Mandrioli2018} or to
machine learning~\cite{Huang2017,Sun2019,Wang2018}.
\end{itemize}

\paragraph{Main Part.}

The main part should offer one or two formal methods of different
nature, e.g. a ``model-oriented'' and a ``property-oriented'' one,
cf.~\cite{wing} for further discussion of this classification; in
order to demonstrate the `universality' of formal methods, it would
appear useful to draw examples from different domains.

The following topics (listed in no particular order) should be covered:
\begin{itemize}
\item
Modelling: going from the informal to the formal; traceability; validation of models.
\item
Language design: explaining how the language of a formal method is designed for specific purposes (what are essentials necessary for expressivity, what is syntactic sugar easing the life of the specifier?).
\item
Semantics: presenting just the essentials -- this needs to be one
topic among many rather than the dominating one, as happens too often
in current practice.
\item
Software engineering context: demonstrating that formal methods are applicable throughout the whole software lifecycle, e.g., in analysing designs, in software verification, testing from formal models.
\item
Method: systematically using tools to illustrate the `method' aspect.
\item
Application domains: illustrate the reach of formal methods by selecting examples from different application domains. Safety, security, human-computer interaction, e-contracts, and non-computer areas (biological systems, ecology, chemistry) are some possible examples.

Traditionally, formal methods teaching advocates the use of formal
methods for safety-critical systems. Formal methods are of course
super-important for those systems, but experience in class (and
otherwise) suggests that this does not inspire and is almost
counterproductive: most students do not foresee themselves designing
the quite narrow range of safety-critical systems we tend to use as
example (airplanes, cars, medical devices, etc.); focusing almost
exclusively on safety-critical systems can actually be
counterproductive as it (can be perceived to) send signals that formal
methods are only usable for such systems.

As cybersecurity failures are much in the news, we might look at these
and see how formal methods might have found these (e.g.\ Heartbleed),
or are being used (e.g.\ Chromium), as a way of emphasising the
mainstream utility of formal methods.
\end{itemize}

\paragraph{Conclusion -- reflection on formal methods.}

We present below some items of reflective nature that ought to be
addressed at the end of a formal methods \module.

\begin{itemize}

\item
General limitations: what formal methods can offer, what formal
methods cannot deliver, e.g., based on Levenson's provocative article
``Are You Sure Your Software Will Not Kill Anyone?''~\cite{levenson}.

\item
Scalability: why formal methods work on toy examples but their
application might become impossible for technical reasons when it
comes to real life challenges, see,
e.g.,~\cite{DBLP:journals/eceasst/RoggenbachMSTN12} and~\cite{james}.
\cite{DBLP:journals/eceasst/RoggenbachMSTN12}~shows a formal methods
in its early stages, where it can barely verify a toy example;
\cite{james}~shows how, after two further years of research, with the
help of abstractions it is possible to verify a real world example
with the very same approach.

\item
Costs/benefits: what the cost and financial benefits of formal methods
are~\cite{barnes11}. The key insight ``Formal methods are surprisingly
feasible for mainstream software development and give good return on
investment.'' from Newcombe et al.~\cite{Newcombeetal2015a} and Amazon's ``We
can now use automated reasoning to provide inexpensive and provable
assurance to customers'' from J.\ {Backes} et al.~\cite{8880058}
are probably a `must have'!

\item
Acceptance: current uptake of formal methods in industry and reasons for the low acceptance.
\item
Current trends: where one expects the field of formal methods to be in, say, a decade.
\end{itemize}
Each lecturer will have her/his own subjective view concerning the
above list of topics. Probably they offer a good point for discussion
with students. The systematic element underlying them is that they
ought to be addressed at the end of a formal methods \module.

\paragraph{Learning outcomes.}
Such a \module would provide the learning outcomes that students
\begin{itemize}
\item understand the thinking behind formal methods and how it differs
  from ad-hoc programming;
\item
are fluent in the application of one or two formal methods to academic examples;
\item
are able to estimate the potential of formal methods with concrete challenges;
\item
are able to critically compare different formal approaches and choose
the most appropriate for a given, specific application.
\end{itemize}

\section{How to assess our teaching efforts?} \label{sec:efforts}

Having introduced changes to teaching, it is important to assess if
they have been successful. In this section, we collect a number of
ideas as to how this could be done. In the order in which they were
contributed, we present a number of personal statements.


\paragraph{\MF.} The obvious measurement is to compare
exam results year after year, assuming that the same person teaches the
course before and after any changes are made. We are working towards
making some changes to our \module that we could compare against the
previous years' results. However, it is also important to survey the
students before and after the course as well as during the lab
sessions to really understand how they are progressing and how
effective the notes, teaching and lab sessions are in improving their
formal methods expertise.

\paragraph{\CD.}
As a teacher on a Degree Apprenticeship programme, I think one such
method of assessing our own teaching methods, is to actually assess
the students' level of understanding by getting them to apply formal
methods in their workplace: students on our programme are employed.
Often, when we teach formal methods, our students have never seen them
before. We tasked our students with producing a work-based portfolio
where they have to apply discrete mathematics to their
workplace. Whilst some students struggle with the task, for most of
them it becomes apparent how beneficial it is in the workplace.
Sometimes it even highlights issues with the existing systems logic.
In my opinion this is the best outcome and therefore would demonstrate
that we have been teaching successfully.


\paragraph{\SK.}
The formal methods community ought to reflect on what it wants to
achieve in teaching. Ultimately, there is no use in being able to
enumerate different formal methods and just being able to use them if
you don't see any reason to do so. Rather, I am in favour of indeed
trying to change (and measure/evaluate) students' opinions and
attitudes.

Employing a formal approach to software engineering is all about the
resulting quality of the product. Thus, a formal methods course needs
to change students' perceptions about software as a product that is
used in different applications and situations -- eventually, even in
safety critical ones. Nobody would cross a bridge that seems like it
might collapse. At the same time, delivering software that is known to
cease working under certain conditions has become quite accepted. Once
students gain an awareness and consciousness for quality aspects of
software, formal methods (and the effort to use them) will appear more
beneficial.

\paragraph{\MR.}


In my teaching experience, students best learn those topics that they
like to do, that they can try themselves, and that provide them with a
feeling of achievement. For teaching practice in formal methods that
means that we ought to run supporting lab classes. These would offer
meaningful examples on which students can successfully apply a formal
method, or explore why some specific formal method fails. In my view,
lab tasks would be well-designed if, say, 80\% of the students can
solve them, i.e., offer them a sense of achievement.

The other objective would be to educate the majority of computer
science students in such a way that they are capable of applying
formal methods in their future careers in industry. This could be
evaluated by looking at dissertations: do the majority of them report
on the application of formal methods when the project concerns
software development?

\paragraph{\PK.}
One criterion could be the number of students that are interested in
writing their dissertation in the field of formal methods. In
particular, our experience is that while formal methods are not in
high demand with students, the ones who finish our formal methods
\modules usually are willing to gain an expert level of knowledge.
Many students stay interested, once they have developed an appetite
for formal methods.

\paragraph{\AC.}
Assessing the effect of teaching changes in standard formal methods
\modules is a tricky task for a number of reasons:
\begin{enumerate}
  \item classes are normally small; \label{reason-small-classes}
  \item even within the small group there is often a large variety in
    background and interest of the
    students; \label{reason-large-variety}
  \item \label{reason-trendy-drives} although students might be
    interested and even successful in using formal methods, in their
    future research or work goals they are driven by more trendy areas
    and topics, where there is little place for the use of formal
    methods.
\end{enumerate}
Reason~\ref{reason-small-classes} prevents us from collecting enough data
to allow us to produce statistically significant results.
It is therefore more important to informally collect personal opinions from students
through discussions, open-ended questionnaires and interviews, rather than
analysing numerical data such as grades and percentage of successful students.

Reason~\ref{reason-large-variety} requires an initial assessment of
the students to be compared with the final objectives that they
achieve at the and of the course (see \MF's statement
earlier in this section).  A possible form of initial assessment is a
questionnaire to be administered during the very first course class.
The questionnaire should aim at the assessment of
\begin{itemize}
  \item mathematical background;
  \item logical and problem solving skills;
  \item experience with the logic and functional programming paradigms;
  \item knowledge of the software engineering concepts that are central in
    formal methods, such as specification, testing, verification, validation,
    assurance.
  \item knowledge of basic logical and set-theoretic concepts such as syntax,
    semantics, theorem, proof, function and more specific computability concepts
    such as decidability, enumerability, undecidability.
  \item perception of more ``exotic'' formal methods concepts such as system state
    and concurrent system.
\end{itemize}

Due to Reason~\ref{reason-trendy-drives}, looking at dissertations or
careers of former students does not really provide a measure of the
achievement of learning objective. In fact, students' pragmatics in
looking for a thesis topic or choosing their professional career may
clash with their academic interests.

\subsection{Summarizing the ideas}

Assessment is often an exercise of producing numbers that can be
compared over several academic years. Here, different criteria have
been suggested and discussed, including:
\begin{itemize}
\item
exam results of a particular \module;
\item
number of dissertations in which formal methods are applied; and
\item
number of dissertations in the area of formal methods.
\end{itemize}
\AC provided arguments why one should look at such numbers
with care.

For teaching a formal method it has been suggested to closely survey
students during the \module (\MF), and to design lab classes
with `guaranteed success', i.e., which are barely contributing to a
differentiation between students in form of marks (\MR).

A slightly deeper looking approach would be to look at students'
opinions and attitudes and see how they change over time (\SK).

\section{Conclusion and outlook} \label{sec:conclusion}


In this white paper, we have analysed why formal methods are seldom
prominently included in computer science and software engineering
curricula. One often heard reason for this is that they fail to
attract students. However, we believe that students often just have
misconceptions about formal methods. Also, the `coolness factor' of
formal methods is low. Finally, formal methods are not visibly used by
industry. It is a myth that formal methods teaching on a basic level
would require a particularly strong mathematical background. We
provided a number of ideas on how to make formal methods more
attractive to students and gave examples of the uptake of formal
methods in industry beyond the critical systems sector.

In the spirit of the workshop ``Formal Methods -- Fun for Everybody'',
this paper has collected a number of `sparkling ideas' that aim at
improving the situation summarised above. We grouped such ideas into
four categories, namely individual teaching delivery,
cf.\ \cref{sec:fun}, making formal methods visible throughout the
syllabus, cf.\ \cref{sec:visibility}, the proposal of a compulsory
formal methods \module, cf.\ \cref{sec:syllabus}, and ideas about how
to measure the effect of teaching changes, cf.\ \cref{sec:efforts}.











With this white paper a start has been made to make formal method
teaching more popular. The ideas and arguments presented are ready to
be picked up in order to improve existing \modules, to design new
\modules, and to make formal methods more prominent in academic
curricula. The participants of the 2019 workshop were enthusiastic
about this topic, and we hope to have shared some of this enthusiasm
with the reader. Let's turn this into a wider movement!

\bibliographystyle{alpha}
\bibliography{doc}

\end{document}